\documentclass[preprint,aps,prc,showpacs,nofootinbib]{revtex4}%
\usepackage{epsfig}
\usepackage{amsmath,graphicx}
\usepackage{amsmath}
\usepackage{graphicx}
\usepackage{amsfonts}
\usepackage{amssymb}%
\providecommand{\U}[1]{\protect\rule{.1in}{.1in}}
\providecommand{\U}[1]{\protect\rule{.1in}{.1in}}

\newcommand\ba{\begin{eqnarray}}
\newcommand\ea{\end{eqnarray}}

\begin{document}
\title{Complete structure dependent analysis of the decay $P\rightarrow l^{+}l^{-}$ }

\author{A. E. Dorokhov$^{1}$,  M. A. Ivanov$^{1}$, S. G. Kovalenko$^{2}$}
\affiliation{\textit{$^{1}$JINR-BLTP, 141980 Dubna, Moscow region, Russian Federation\\
$^{2}$ Centro de Estudios Subat\'omicos(CES),
Universidad T\'ecnica Federico Santa Mar\'\i a, \\
Casilla 110-V, Valpara\'\i so, Chile}}

\date{\today}

\begin{abstract}
We use the Mellin-Barnes representation in order to improve the theoretical
estimate of mass corrections to the width of light pseudoscalar meson decays
into a lepton pair, $P\rightarrow l^{+}l^{-}$ . The full resummation of the
terms $\left(  M^{2}/\Lambda^{2}\right)  ^{n},$ $\left(  m^{2}/M^{2}\right)
^{n}$and $\left(  m^{2}/\Lambda^{2}\right)  ^{n}$ to the decay amplitude is
performed, where $m$ is the lepton mass, $M$ is the meson mass and
$\Lambda\approx m_{\rho}$ is the characteristic scale of the $P\rightarrow
\gamma^{\ast}\gamma^{\ast}$ form factor. The total effect of mass corrections
is quite important for $\eta(\eta^{\prime})$ decays. We also comment on the
estimation of the hadronic light-by-light scattering contribution to
the muon anomalous magnetic moment in the chiral perturbation theory.
\end{abstract}
\maketitle


\section{Introduction}

The theoretical study of the $\pi^{0}$ and $\eta\left(  \eta^{\prime}\right)
$ mesons decaying into lepton pairs and the comparison with the experimental
rates offers an important low-energy test of the standard model. The situation
with these decays became more pressing after recent KTeV E799-II experiment at
Fermilab \cite{Abouzaid:2007md} in which the pion decay into an electron-positron pair was measured
with high accuracy using the $K_{L}\rightarrow3\pi$ process as a source of
tagged neutral pions
\begin{equation}
B_{\mathrm{no-rad}}^{\mathrm{KTeV}}\left(  \pi^{0}\rightarrow e^{+}%
e^{-}\right)  =\left(  7.49\pm0.29\pm0.25\right)  \cdot10^{-8}.\label{KTeV}%
\end{equation}
The theory for the decay of pseudoscalar mesons to a lepton pair is known for
decades \cite{Drell59,Berman60,Bergstrom:1982zq,Efimov:1981vh}. The main
limitation for realistic prediction of these processes comes from the large
distance contributions of the strong sector of the standard model where the
perturbative QCD theory does not work. However, it was shown in
\cite{Dorokhov:2007bd} that theoretical uncertainty can be significantly
reduced by using CELLO and CLEO data \cite{Behrend:1990sr,Gronberg:1997fj} and
QCD\ constraints on the transition form factors $P\rightarrow\gamma^{\ast
}\gamma^{\ast}$. As a result, the standard model prediction gives
\cite{Dorokhov:2007bd}
\begin{equation}
B^{\mathrm{Theor}}\left(  \pi^{0}\rightarrow e^{+}e^{-}\right)  =\left(
6.2\pm0.1\right)  \cdot10^{-8},\label{Bth}%
\end{equation}
which is $3.3\sigma$ below the KTeV result (\ref{KTeV}). It is extremely
important to trace possible sources of the discrepancy between the experiment
and theory. There are a number of possibilities: (i) problems with (statistic)
experiment procession, (ii) inclusion of QED radiation corrections by KTeV is
incomplete, (iii) unaccounted mass corrections are important, and (iv) effects
of new physics. At the moment, the last possibility was reinvestigated.

In \cite{Dorokhov:2008qn}, the contribution of QED radiative corrections to
the $\pi^{0}\rightarrow e^{+}e^{-}$ decay, which must be taken into account
when comparing the theoretical prediction (\ref{Bth}) with the experimental
data, was revised. In comparison with earlier studies \cite{Bergstrom:1982wk}
the main progress made in \cite{Dorokhov:2008qn} consists in detailed analysis
of the $\gamma^{\ast} \gamma^{\ast}\rightarrow e^{+}e^{-}$ subprocess and
revealing the dynamics of long and short distances. Occasionally, the final
result agrees well with earlier prediction based on calculations
\cite{Bergstrom:1982wk} and, thus, the KTeV analysis of radiative corrections
is confirmed.

There are quite few attempts in the literature  to explain the excess of the
experimental data on the $\pi^{0}\rightarrow e^{+}e^{-}$ decay over the standard
model predictions as a manifestation of physics beyond the Standard Model. In
Ref. \cite{Kahn:2007ru}, it was shown that this excess could be explained
within the currently popular model of light dark matter involving a low mass
($\sim10$ MeV) vector bosons $U_{\mu}$ which presumably couples to the
axial-vector currents of quarks and leptons. Another possibility was proposed
in Ref. \cite{Chang:2008np,McKeen:2008gd} in interpreting the same experimental
effect as the contribution of the light CP-odd Higgs boson appearing in the
next-to-minimal supersymmetric Standard Model. However, there are other
extensions of the Standard Model deserving studies in this context. In particular,
supersymmetric models with R-parity violation and leptoquark models suggest
contributions to this process which under certain circumstances may be
significant. Specific examples of such contributions are related to the
exchange by the t-squark and the $SU_{2L}$ singlet leptoquarks with the
couplings to u,d-quarks and electrons not stringently constrained from other
known processes \cite{Chemtob:2004xr,Barbier:2004ez}.

In the present paper, we focus on the mass corrections to the width of light
pseudoscalar meson decays into a lepton pair, $P\rightarrow l^{+}l^{-}$. We
show that these corrections are under theoretical control and do not help
resolving the problem of disagreement of the KTeV data on the $\pi
^{0}\rightarrow e^{+}e^{-}$ decay with the standard model prediction. However,
the mass corrections are quite important for realistic predictions for
$\eta\left(  \eta^{\prime}\right)  $ decays to a lepton pair. \begin{figure}[th]
\includegraphics[width=5cm]{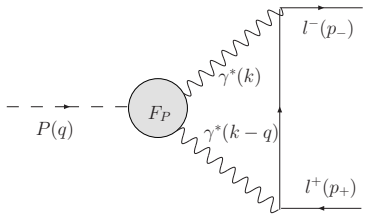}\caption{Triangle diagram for the
$P\rightarrow l^{+}l^{-}$ process with the pseudoscalar meson form factor
$P\rightarrow\gamma^{\ast}\gamma^{\ast}$ in the vertex.}%
\label{fig:triangle}%
\end{figure}

In the lowest order of QED perturbation theory, the decay of the neutral
meson, $P(q)\rightarrow l^{-}(p_{-})+l^{+}(p_{+}),\quad q^{2}=M^{2},\quad
p_{\pm}^{2}=m^{2},$ ($M$ meson mass, $m$ lepton mass) is described by the
one-loop Feynman amplitude (Fig. \ref{fig:triangle}) corresponding to the
conversion of the neutral meson through two virtual photons into a lepton
pair. The normalized branching ratio is given by
\cite{Drell59,Berman60,Efimov:1981vh,Bergstrom:1982zq}
\begin{equation}
R_{0}(P\rightarrow l^{+}l^{-})=\frac{B_{0}\left(  P\rightarrow l^{+}%
l^{-}\right)  }{B\left(  P\rightarrow\gamma\gamma\right)  }=2\beta\left(
M^{2}\right)  \left(  \frac{\alpha m}{\pi M}\right)  ^{2}|\mathcal{A}\left(
M^{2}\right)  |^{2}, \label{Bpi}%
\end{equation}
where $\beta\left(  q^{2}\right)  =\sqrt{1-4m^{2}/q^{2}}$ and the reduced
amplitude is
\begin{equation}
\mathcal{A}\left(  q^{2}\right)  =\frac{2}{q^{2}}\int\frac{d^{4}k}{i\pi^{2}%
}\frac{(qk)^{2}-q^{2}k^{2}}{(k^{2}+i\epsilon)\left[  (q-k)^{2}+i\epsilon
\right]  \left[  (p_{-}-k)^{2}-m^{2}+i\epsilon\right]  }F_{P\gamma^{\ast
}\gamma^{\ast}}(-k^{2},-(q-k)^{2}), \label{Rq}%
\end{equation}
with the transition form factor $F_{P\gamma^{\ast}\gamma^{\ast}}(-k^{2}%
,-q^{2})$ being normalized as $F_{P\gamma^{\ast}\gamma^{\ast}}(0,0)=1$. It is
convenient to introduce the mass ratio parameters:%
\begin{equation}
x=\left(  m/\Lambda\right)  ^{2},\quad u=\left(  m/M\right)  ^{2},\quad
z=x/u=\left(  M/\Lambda\right)  ^{2}, \label{xuz}%
\end{equation}
where $\Lambda\sim M_{\rho}$ is a parameter characterizing the transition form
factor. In the physically interesting cases the parameter $x$ is very small:
$(m_{e}/\Lambda)^{2}\lesssim10^{-7},(m_{\mu}/\Lambda)^{2}\lesssim10^{-2}$. At
the same time, we consider the amplitude in the leading order in the
fine coupling constant $\alpha\sim10^{-2}$. Thus, in the following it is
reasonable to neglect the $x$ power dependence of the amplitude keeping only the
dependence on the $u$ and $z$ variables \footnote{Partially, the dependence of
the amplitude $\mathcal{A}\left(  q^{2}\right)  $ on the $x$ parameter was
taken into account in \cite{Dorokhov:2008cd}.}. The aim of this work is to improve the
previous calculations of the amplitude $\mathcal{A}\left(  M^{2}\right)  $ of
the $P\rightarrow l^{+}l^{-}$ decay by taking into account all order mass corrections.

\section{Resummation of power corrections to the amplitude}

We evaluate the amplitude $\mathcal{A}\left(  q^{2}\right)  $ following the
way used in \cite{Efimov:1981vh,Dorokhov:2007bd,Dorokhov:2008cd}. Let us
transform the integral in (\ref{Rq}) to the Euclidean metric $k_{0}\rightarrow
ik_{4}$. The corresponding integral is convergent due to decreasing of
$F_{P\gamma^{\ast}\gamma^{\ast}}(k^{2},(q-k)^{2})$ in the Euclidean region. It
is convenient to introduce the modified Mellin-Barnes transformation for the
meson form factor%
\begin{equation}
F_{P\gamma^{\ast}\gamma^{\ast}}(k^{2},(q-k)^{2})=\frac{1}{\left(  2\pi
i\right)  ^{2}}\int_{\sigma+iR^{2}}dz_{1}dz_{2}\Phi\left(  z_{1},z_{2}\right)
\Gamma\left(  z_{1}\right)  \Gamma\left(  z_{2}\right)  \left(  \frac
{\Lambda^{2}}{k^{2}}\right)  ^{z_{1}}\left(  \frac{\Lambda^{2}}{\left(
k-q\right)  ^{2}}\right)  ^{z_{2}},\label{FM}%
\end{equation}
where $\Lambda$ is the characteristic scale for the form factor, the vector
$\sigma=\left(  \sigma_{1},\sigma_{2}\right)  \in\mathbb{R}^{2},$ and
$\Phi\left(  z_{1},z_{2}\right)  $ is the inverse Mellin-Barnes transform of
the form factor
\begin{equation}
\Phi\left(  z_{1},z_{2}\right)  =\frac{1}{\Gamma\left(  z_{1}\right)
\Gamma\left(  z_{2}\right)  }\int_{0}^{\infty}dt_{1}\int_{0}^{\infty}%
dt_{2}t_{1}^{z_{1}-1}t_{2}^{z_{2}-1}F_{P\gamma^{\ast}\gamma^{\ast}}\left(
t_{1},t_{2}\right)  \label{PhiM}%
\end{equation}
which is an analytical function of $z_{1}$ and $z_{2}$. Introducing Feynman
parameterization in (\ref{Rq}) we convert the $k-$loop integral into the
integrals in Feynman parameters which can be expressed in terms of $\Gamma
$-functions. Then we obtain the following Mellin-Barnes representation for
the amplitude \cite{Dorokhov:2008cd}%
\begin{align}
\mathcal{A}\left(  q^{2}\right)   &  =-\int_{\sigma+i\mathbb{R}^{3}}%
\frac{dz_{1}dz_{2}dz_{3}}{\left(  2\pi i\right)  ^{3}}x^{-z_{1}-z_{2}%
}u^{-z_{3}}\Phi\left(  z_{1},z_{2}\right)  \left[  3-\left(  2+\frac{1}%
{2u}\right)  \left(  z_{1}+z_{2}+z_{3}\right)  \right]  \label{A0}\\
&  \frac{\Gamma\left(  -z_{3}\right)  \Gamma\left(  1+z_{1}+z_{3}\right)
\Gamma\left(  1+z_{2}+z_{3}\right)  \Gamma\left(  z_{1}+z_{2}+z_{3}\right)
\Gamma\left(  1-2\left(  z_{1}+z_{2}+z_{3}\right)  \right)  }{z_{1}z_{2}%
\Gamma\left(  3-z_{1}-z_{2}\right)  },\nonumber
\end{align}
with $\sigma$ in the 3-dimensional region in space of real parts of $z_{i}$
chosen so that the integration path $\sigma+i\mathbb{R}^{3}$ does not
intersect the $\Gamma$-function singularities. Finally, we are able to expand
the integral over the $x$, $u$ and $z$ mass ratios by closing the contours of
integration in the appropriate manner and summing up the obtained series.
Neglecting the dependence of the amplitude $\mathcal{A}\left(  q^{2}\right)  $ on
the powers of small $x$ parameter (but keeping $\ln x$) we arrive at the
following representation:
\begin{align}
\mathcal{A}\left(  M^{2}\right)   &  =\frac{1}{\beta\left(  u\right)  }\left[
\frac{1}{4}\ln^{2}\left(  y\left(  u\right)  \right)  +\frac{\pi^{2}}%
{12}+\mathrm{Li}_{2}\left(  -y\left(  u\right)  \right)  \right]
\label{ReA}\\
&  +i\frac{\pi}{2\beta\left(  u\right)  }\ln\left(  y\left(  u\right)
\right)  \nonumber\\
&  +\frac{3}{2}\ln\left(  x\right)  -\frac{5}{4}+\frac{3}{2}\int_{0}^{\infty
}dt\ln\left(  \frac{t}{\Lambda^{2}}\right)  F_{P\gamma^{\ast}\gamma^{\ast}%
}^{\left(  1,0\right)  }\left(  t,t\right)  \nonumber\\
&  +A_{z},\nonumber\label{ImA}%
\end{align}
where
\[
y\left(  q^{2}\right)  =\frac{1-\beta\left(  q^{2}\right)  }{1+\beta\left(
q^{2}\right)  },
\]
and the correction to $\mathcal{A}\left(  M^{2}\right)  $ is expressed as%
\begin{align}
&  A_{z}=\left(  \ln x+\frac{3}{2}\right)  \int_{0}^{1}\frac{ds}{s}\left(
1-s\right)  ^{2}\left[  1-F_{P\gamma^{\ast}\gamma^{\ast}}\left(  -sz,0\right)
\right]  -\frac{z}{3}F_{P\gamma^{\ast}\gamma^{\ast}}^{\left(  1,0\right)
}\left(  0,0\right)  \label{Az}\\
&  -\frac{z}{6}\int_{0}^{\infty}\frac{dt}{t^{2}}\left[  F_{P\gamma^{\ast
}\gamma^{\ast}}(t,t)-F_{P\gamma^{\ast}\gamma^{\ast}}(0,t)-tF_{P\gamma^{\ast
}\gamma^{\ast}}^{\left(  1,0\right)  }\left(  0,t\right)  \right]  \nonumber\\
&  +\int_{0}^{1}\frac{ds}{s}\left(  1-s\right)  ^{2}\int_{0}^{1}\frac
{dy}{\left(  1-y\right)  y}\nonumber\\
&  \times\left[  1-2y+y^{3}+\frac{3}{2}yF_{P\gamma^{\ast}\gamma^{\ast}}\left(
-sz,0\right)  -\left(  1-\frac{1}{2}y+y^{3}\right)  F_{P\gamma^{\ast}%
\gamma^{\ast}}\left(  -syz,0\right)  \right]  \nonumber\\
&  +\frac{1}{2}\int_{0}^{1}\frac{ds}{s}\left(  1-s\right)  ^{2}\int
_{0}^{\infty}dt\ln(t)\left[  2F_{P\gamma^{\ast}\gamma^{\ast}}^{\left(
1,0\right)  }\left(  t,0\right)  -F_{P\gamma^{\ast}\gamma^{\ast}}^{\left(
1,0\right)  }\left(  t-sz,0\right)  -F_{P\gamma^{\ast}\gamma^{\ast}}^{\left(
1,0\right)  }\left(  t,-sz\right)  \right]  \nonumber\\
&  -\frac{3z}{\pi}\int_{0}^{1}ds\left(  1-s\right)  ^{2}\int_{0}^{\pi/2}dv\cos
v\int_{0}^{\infty}dt\frac{1}{\left(  stz\right)  ^{1/2}}\operatorname{Im}%
F_{P\gamma^{\ast}\gamma^{\ast}}^{\left(  1,0\right)  }\left(  t-zs+2i\left(
stz\right)  ^{1/2}\cos v,t\right)  .\nonumber
\end{align}
In the above expressions $F_{P\gamma^{\ast}\gamma^{\ast}}^{\left(
\alpha,\beta\right)  }(s,t)$ denotes the derivatives of an order of $\alpha$
and $\beta$ in the corresponding arguments of the form factor.

The first and second lines in (\ref{ReA}) are the structure independent parts
of the amplitude. These expressions agree with obtained earlier results. In
particular, they agree with the results of the dispersion approach of
\cite{Bergstrom:1983ay} and the chiral perturbation theory
\cite{D'Ambrosio:1986ze,Savage:1992ac}. The derivation of the amplitude by
using the dispersion approach tacitly assumes that the imaginary part of the
off-shell amplitude $\mathcal{A}(q^{2})$ is the second line of (\ref{ReA})
with $M^{2}$ substituted by $q^{2}$. Thus, it takes into account only the
structure independent part (dependence on the parameter $u$) and is
insensitive to the details of the transition form factor. The same result
appears in the framework of the leading order of the chiral perturbation
theory \cite{D'Ambrosio:1986ze,Savage:1992ac}, because at this order one also
does not take into account the form factor. The integral in the second line of
(\ref{ReA}) was estimated in \cite{Dorokhov:2007bd} by using CELLO and CLEO
data on the meson transition form factors and constraints following from
operator product expansion (OPE) in QCD. Our new result concerns the structure
dependent part of the amplitude $A_{z}$ as the combination of the integrals of
the meson transition form factor over both the spacelike and timelike regions.

If we retain only the terms linear in $z$ we get
\begin{align}
A_{z}^{(1)}  &  =-\frac{z}{6}\left\{  \left(  -2\ln x+\frac{14}{3}\right)
F_{P\gamma^{\ast}\gamma^{\ast}}^{\left(  1,0\right)  }\left(  0,0\right)
+3\int_{0}^{\infty}dtF_{P\gamma^{\ast}\gamma^{\ast}}^{\left(  1,1\right)
}\left(  t,t\right)  \right. \label{Delta}\\
&  -\int_{0}^{\infty}dt\ln(t)\left[  F_{P\gamma^{\ast}\gamma^{\ast}}^{\left(
1,1\right)  }\left(  t,0\right)  +F_{P\gamma^{\ast}\gamma^{\ast}}^{\left(
2,0\right)  }\left(  t,0\right)  \right] \nonumber\\
&  \left.  +\int_{0}^{\infty}\frac{dt}{t^{2}}\left[  F_{P\gamma^{\ast}%
\gamma^{\ast}}(t,t)-F_{P\gamma^{\ast}\gamma^{\ast}}(0,t)-tF_{P\gamma^{\ast
}\gamma^{\ast}}^{\left(  1,0\right)  }\left(  0,t\right)  \right]  \right\}
.\nonumber
\end{align}

In order to estimate the correction $A_{z}$ to the amplitude related to the
meson mass, we consider the simplest parameterization of the transition form
factor given by the naive vector meson dominance model
\begin{equation}
F_{P\gamma^{\ast}\gamma^{\ast}}^{\mathrm{VMD}}\left(  s,t\right)  =\frac
{M_{V}^{4}}{\left(  M_{V}^{2}+s\right)  \left(  M_{V}^{2}+t\right)  }.
\label{VMD}
\end{equation}
In that case, $z=\left(  M/M_{V}\right)  ^{2}$ and the correction is%
\begin{align}
A_{z}^{\mathrm{VMD}}  &  =\ln\left(  x\right)  \left(  \frac{\left(
1-z\right)  ^{2}}{z^{2}}\ln(1-z)-\frac{3}{2}+\frac{1}{z}\right)  -\frac
{2z-1}{z^{2}}\mathrm{Li}_{2}\left(  z\right)  +\label{Avmd}\\
&  +\frac{\left(  1-z\right)  }{2z^{2}}\ln\left(  1-z\right)  \left[  \left(
1-z\right)  \ln\left(  1-z\right)  +6z-4\right]  +\frac{z}{6}+\frac{3}%
{2}-\nonumber\\
&  -\frac{3}{4z^{2}}\left\{  -\mathrm{arctg}^{2}\left[  \frac{\sqrt{4z-z^{2}%
}\left(  2-z\right)  }{2-4z+z^{2}}\right]  \right. \nonumber\\
&  \left.  +\sqrt{\frac{z}{4-z}}\left(  8-6z+z^{2}\right)  \mathrm{arctg}%
\left[  \frac{\sqrt{4z-z^{2}}\left(  2-z\right)  }{2-4z+z^{2}}\right]
\right\} \nonumber
\end{align}
which in the linear approximation ($z<<1$) reduces to
\begin{equation}
A_{z}^{(1),\mathrm{VMD}}=\frac{z}{6}\left(  -2\ln x+\frac{5}{3}\right)  .
\end{equation}

Taking $M_{V}=770$ MeV the resulting branchings are given in the Table and
compared with existing experimental data. The so-called unitary bound appears
if in (\ref{Bpi}) only the imaginary part of the amplitude which is model
independent is taken into account. The CLEO bound corresponds to the estimate
of the real part of the amplitude basing on the CELLO and CLEO data on the
meson transition from factors. The fourth column of the Table contains the
predictions where in addition the constraint from OPE QCD on the transition
form factor for arbitrary photon virtualities is taken into
account\cite{Dorokhov:2007bd}. \begin{table}[ptb]
\caption[Results]{Values of the branchings $B\left(  P\rightarrow l^{+}%
l^{-}\right)  $ obtained in our approach and compared with the available
experimental results. }%
\label{table2}
\begin{tabular}
[c]{|c|c|c|c|c|c|}\hline
$R_{0}$ & Unitary bound & CLEO bound & CLEO+OPE & This work & Experiment\\
&  &  &  &  & \\\hline
$R_{0}\left(  \pi^{0}\rightarrow e^{+}e^{-}\right)  \times10^{8}$ & $\geq4.69$
& $\geq5.85\pm0.03$ & $6.23\pm0.12$ & $6.26$ & $7.49\pm0.38$
\cite{Abouzaid:2007md}\\\hline
$R_{0}\left(  \eta\rightarrow\mu^{+}\mu^{-}\right)  \times10^{6}$ & $\geq4.36$
& $\leq6.23\pm0.12$ & $5.12\pm0.27$ & $4.64$ & $5.8\pm0.8$
\cite{Yao:2006px,Abegg:1994wx}\\\hline
$R_{0}\left(  \eta\rightarrow e^{+}e^{-}\right)  \times10^{9}$ & $\geq1.78$ &
$\geq4.33\pm0.02$ & $4.60\pm0.09$ & $5.24$ & $\leq2.7\cdot10^{4}$
\cite{Berlowski:2008zz}\\\hline
$R_{0}\left(  \eta^{\prime}\rightarrow\mu^{+}\mu^{-}\right)  \times10^{7}$ &
$\geq1.35$ & $\leq1.44\pm0.01$ & $1.364\pm0.010$ & $1.30$ & \\\hline
$R_{0}\left(  \eta^{\prime}\rightarrow e^{+}e^{-}\right)  \times10^{10}$ &
$\geq0.36$ & $\geq1.121\pm0.004$ & $1.178\pm0.014$ & $1.86$ & \\\hline
\end{tabular}
\end{table}

As expected, a visible change occurs only for $\eta\left(  \eta^{\prime
}\right)  $ meson decays. It is interesting that for $\eta$ decay to muons the
mass correction shifts the theoretical prediction in the direction to the
unitary bound and thus opposite to the experimental result \cite{Abegg:1994wx}%
. This is because the real part of the amplitude for this process taken at the
physical point $q^{2}=M_{\eta}^{2}$ remains negative and a positive shift due
to mass correction reduces the absolute value of the real part of the
amplitude $\left\vert \operatorname{Re}\mathcal{A}\left(  M_{\eta}^{2}\right)
\right\vert $. In this situation, new measurements of $\eta\rightarrow\mu
^{+}\mu^{-}$ would be very desirable.

For $\eta^{\prime}$ decays there appear new thresholds in addition to
the two-photon one. In general, this violates the so-called unitary bound because the
correction $A_z$ gains an imaginary part at $z>1$
\begin{equation}
\Delta\operatorname{Im}\mathcal{A}   =-\frac{\pi}{\beta}\left(  1-\frac{1}{z}\right)
^{2}\ln\left(  \frac{1+\beta}{1-\beta}\right) \Theta(z-1) .
\end{equation}
 As it is
seen from the table, this happens for the $\eta^{\prime}\rightarrow\mu\mu$
channel. Nevertheless, it turns out that the predictions for this channel are
quite accurate. We checked that more sophisticated models for the transition
form factor (the generalized VMD \cite{Knecht:2001xc} or the effect of meson
mixing \cite{Silagadze:2006rt}) do not significantly change the results of
the Table.

\section{Hadronic light-by-light contribution to muon g-2 in chiral
perturbation theory}

In \cite{RamseyMusolf:2002cy}, the hadronic light-by-light contribution to muon
$g-2$, which is enhanced by large logarithms (LL) and a factor of $N_{c}$
$\left(  \sim\mathcal{O}\left(  \alpha^{3}N_{c}m_{\mu}^{2}\right)  LL\right)
$, was estimated in the chiral perturbation theory as%
\begin{equation}
a_{\mu}^{\mathrm{LbL,hadr}}=a_{\mu,\mathrm{LL}}^{\mathrm{LbL,hadr}}%
+a_{\mu,\mathrm{pionloop}}^{\mathrm{LbL,hadr}}+\widetilde{C}, \label{aLbLlog}%
\end{equation}
where%
\[
a_{\mu,\mathrm{LL}}^{\mathrm{LbL,hadr}}=\frac{3}{16}\left(  \frac{\alpha}{\pi
}\right)  ^{3}\left(  \frac{m_{\mu}}{F_{\pi}}\right)  ^{2}\left(  \frac{N_{c}%
}{3\pi}\right)  ^{2}\left\{  \frac{1}{4}\ln^{2}\left(  \frac{m_{\mu}^{2}%
}{\Lambda^{2}}\right)  -\frac{1}{2}\ln\left(  \frac{m_{\mu}^{2}}{\Lambda^{2}%
}\right)  \left[  -f\left(  \frac{m_{\mu}^{2}}{m_{\pi}^{2}}\right)  +\frac
{1}{2}+\frac{\chi\left(  \Lambda\right)  }{6}\right]  \right\}  ,
\]%
\[
f\left(  y\right)  =\frac{1}{6}y^{2}\ln y-\frac{1}{6}\left(  2y+13\right)
+\frac{1}{3}\left(  2+y\right)  \sqrt{y\left(  4-y\right)  }\arccos\left(
\frac{\sqrt{y}}{2}\right)  ,
\]
the constant $\widetilde{C}$ absorbs subdominant structure dependent
contributions. The low energy constant $\chi\left(  \Lambda\right)  $ is
related to the amplitude for pion decay into an electron-positron pair
by\footnote{The logarithmically enhanced hadronic light-by-light contributions
to $a_{\mu}$ are renormalization scheme independent. However, the values
$\chi\left(  \Lambda\right)  $, $f\left(  y\right)  $ and the constant
$(1/2)$ appearing in the renormalization group equation depend on the
choice of a scheme. Thus, the constant $\chi\left(  \Lambda\right)  $ defined in
\cite{RamseyMusolf:2002cy} is related to the corresponding constant $\chi^{\ast
}\left(  m_{\rho}\right)  $\ in \cite{Knecht:1999gb} by relation $\chi\left(
\Lambda\right)  =-4\left[  \chi^{\ast}\left(  m_{\rho}\right)  -3ln\left(
\frac{m_{\rho}}{\Lambda}\right)  +1\right]  $.}%
\begin{align}
\chi\left(  \Lambda\right)   &  =2\left(  3\ln\left(  \frac{m_{e}^{2}}%
{\Lambda^{2}}\right)  -2\mathcal{A}\left(  0\right)  -7\right) \label{ChiL}\\
&  =-9-6\int_{0}^{\infty}dt\ln\left(  \frac{t}{\Lambda^{2}}\right)
F_{P\gamma^{\ast}\gamma^{\ast}}^{\left(  1,0\right)  }\left(  t,t\right)
.\nonumber
\end{align}
Taking $\Lambda=1$ GeV we obtain for the logarithmic enhanced term
\begin{equation}
a_{\mu,\mathrm{LL}}^{\mathrm{LbL,hadr}}=\left(  3.4\pm2.0\right)
\times10^{-10}, \label{aLL}%
\end{equation}
where uncertainty arises from determination of the integral in (\ref{ChiL}).
The previous estimate was $a_{\mu,\mathrm{LL}}^{\mathrm{LbL,hadr}}=\left(
5.7_{-6.0}^{+5.0}\right)  \times10^{-10}$\cite{RamseyMusolf:2002cy}. Better
accuracy in (\ref{aLL}) is due to higher quality of determination of
$\chi\left(  1\mathrm{GeV}\right)  =-17.35\pm1.20$ by using CELLO and CLEO
data for the transition from factor $F_{P\gamma^{\ast}\gamma^{\ast}}\left(
t,t\right)  $ instead of using the decay $\eta\rightarrow\mu^{+}\mu^{-}$, as
was done in \cite{RamseyMusolf:2002cy} with the result $\chi\left(
1\mathrm{GeV}\right)  =-14_{-5}^{+4}$.

The second term in (\ref{aLbLlog}) $a_{\mu,\mathrm{pionloop}}%
^{\mathrm{LbL,hadr}}$ being of the order $O\left(  \alpha^{3}N_{c}^{0}\right)
$ arises from the three-loop graphs with a charged pion loop. It was
computed in \cite{Hayakawa:1995ps,Kinoshita:1984it} $a_{\mu,\mathrm{pionloop}%
}^{\mathrm{LbL,hadr}}=-4.46\times10^{-10}.$ Thus, there are large
cancellations between $\ln^{2}$ and $\ln$ terms in $a_{\mu\partial
,\mathrm{log}}^{\mathrm{LbL,hadr}}$ as well as between $a_{\mu,\mathrm{log}%
}^{\mathrm{LbL,hadr}}$ and $a_{\mu,\mathrm{pionloop}}^{\mathrm{LbL,hadr}}$
with final result
\begin{equation}
a_{\mu,\chi PT}^{\mathrm{LbL,hadr}}=\left(  -1.06\pm2.0+3.1\widetilde
{C}\right)  \times10^{-10}.
\end{equation}

The conclusion is that in the case of $P\rightarrow ll$ decays the large
logarithmic terms are dominant and the structure dependent part of the
amplitude is a small correction. That is why the predictions for the
$P\rightarrow ll$ decay branchings are rather stable. This is not the case for
the hadronic light-by-light contribution to muon $g-2$. Due to cancellations,
the nonleading structure dependent terms become crucially important and the
chiral perturbation theory does not provide realistic estimate. It is
necessary to use specific models to provide correct predictions for $a_{\mu
}^{\mathrm{LbL,hadr}}$ (see for review
\cite{Miller:2007kk,Passera:2007fk,Dorokhov:2005ff,Jegerlehner:2009ry}).

\section{Conclusions}

Our main conclusion is that the inclusion of radiative and mass corrections is
unable to reduce the discrepancy between the standard model prediction for the
$\pi^{0}\rightarrow e^{+}e^{-}$ decay rate (\ref{Bth}) and experimental result
(\ref{KTeV}). Mass corrections are important to get a reliable prediction for
branchings of $\eta$-meson to a lepton pair. Moreover, they are small enough to
be sensitive to details of the transition form factor. For $\eta^{\prime}$
decay modes not only mass corrections but also threshold behavior of the
transition form factor are important. Thus, theoretical error for $\eta
^{\prime}$ decays is less under control comparing with $\pi^{0}$ and $\eta$
meson decays. Further independent experiments at KLOE \cite{Bloise:2008zz},
WASAatCOSY \cite{Kupsc:2008zz}, BES III \cite{Li:2009jd} and other facilities
will be crucial for resolution of the problem.

\section{Acknowledgments}

A.E.D. acknowledges partial support from the Scientific School grant
4476.2006.2.
This work is partially supported by the PBCT project ACT-028 ÒCenter of
Subatomic PhysicsÓ.

\end{document}